\documentclass[10pt, aps,prx,superscriptaddress, twocolumn, floatfix,nolongbibliography]{revtex4-2}
\usepackage{graphicx}
\usepackage{xcolor}
\usepackage{textcomp}
\usepackage{siunitx}
\usepackage{bm}
\usepackage[normalem]{ulem}
\usepackage{amsmath, amssymb}
\usepackage{amsfonts}
\usepackage{adjustbox}
\usepackage[utf8]{inputenc}
\usepackage{placeins}

\begin{document}

\title{\large{
Systematic Characterization of Transmon Qubit Stability with Thermal Cycling
}}

\author{Cong Li}
\affiliation{College of Computer Science and Technology, National University of Defense Technology, Changsha 410073, China}
\author{Zhaohua Yang}
\affiliation{College of Computer Science and Technology, National University of Defense Technology, Changsha 410073, China}
\author{Xinfang Zhang}
\affiliation{College of Computer Science and Technology, National University of Defense Technology, Changsha 410073, China}
\author{Zhihao Wu}
\affiliation{College of Computer Science and Technology, National University of Defense Technology, Changsha 410073, China}
\author{Shichuan Xue}
\email{Corresponding author: Shichuanxue@quanta.org.cn}
\affiliation{College of Computer Science and Technology, National University of Defense Technology, Changsha 410073, China}
\author{Mingtang Deng}
\email{Corresponding author: mtdeng@nudt.edu.cn}
\affiliation{College of Computer Science and Technology, National University of Defense Technology, Changsha 410073, China}

\date{\today}

\begin{abstract}
The temporal stability and reproducibility of qubit parameters are critical for the long-term operation and maintenance of superconducting quantum processors. In this work, we present a comprehensive longitudinal characterization of 27 frequency-tunable transmon qubits spanning over one year across four thermal cycles. Our results establish a distinct hierarchy of stability for superconducting hardware. We find that the intrinsic device parameters determining the qubit frequency and the baseline energy relaxation times ($T_1$) exhibit high robustness against thermal stress, characterized by frequency deviations typically confined within 0.5\% and non-degraded coherence baselines. In stark contrast, the environmental variables, specifically the background magnetic flux offsets and the microscopic landscape of two-level system (TLS) defects, undergo a significant stochastic reconfiguration after each cycle. By employing frequency-dependent relaxation spectroscopy and a quantitative metric, the $T_1$ Spectral Topography Fidelity, we demonstrate that thermal cycling acts as a ``hard reset'' for the local defect environment. This process introduces a level of spectral randomization equivalent to thousands of hours of continuous low-temperature evolution. These findings confirm that while the fabrication quality is preserved, the specific noise realization is statistically distinct for each thermal cycle, necessitating automated recalibration strategies for large-scale quantum systems.
\end{abstract}

\maketitle

\section{Introduction}
The transition of superconducting quantum computing from prototype demonstrations to large-scale processors~\cite{Arute2019, Wu2021, kim2023evidence,xue2022variational} marks a pivotal shift in research priorities~\cite{google2023suppressing}. Early research established coherence in isolated devices~\cite{Place2021, Wang2022, Bland2025}, whereas current efforts must address the long-term stability and reliability of integrated systems~\cite{Siddiqi2021, kjaergaard2020, proctor2022measuring,wu2025telegraph}. The operational requirements for maintenance and upgrades subject these systems to repeated thermal cycling. Large thermal swings between room temperature and millikelvin conditions are known to modify material properties, including inducing chip strain~\cite{yelton2025correlated, grabovskij2012strain, Thorbeck2023}, rearranging trapped vortices~\cite{stan2004critical,song2009microwave}, and shifting microscopic defects~\cite{Muller2019,shalibo2010lifetime}. These potential perturbations represent a critical challenge; however, systematic studies quantifying their impact on device parameters remain notably absent.

Previous studies on the effect of thermal cycling on two-level systems (TLSs) have primarily focused on tracking discrete, strongly coupled defects~\cite{bilmes2021quantum} or have used thermal cycling as a tool to remove a strongly coupled TLS~\cite{Gong2019}. In addition to strongly coupled TLSs, a weak TLS bath surrounding a qubit forms a dense spectral background that fundamentally determines the baseline energy relaxation time ($T_1$)~\cite{heath2026localized} and device stability~\cite{Muller2019}. Nevertheless, the influence of thermal cycling on the off-resonant, weak TLS ensemble is less explored. A systematic understanding of how this collective environment responds to thermal cycling, specifically whether its global spectral landscape is preserved or randomized, is still lacking.

This work systematically characterizes the stability of superconducting qubits based on a longitudinal dataset obtained from 27 selected qubits over one year and across four thermal cycles. By shifting the focus from discrete strong defects to the collective spectral topography of the TLS bath, we introduce a quantitative metric, the \textit{$T_1$ Spectral Topography Fidelity} (STF), to evaluate correlations in the qubit relaxation landscape. Our results reveal a clear hierarchy of stability. While the intrinsic hardware parameters that determine qubit frequency remain highly stable, the microscopic environment undergoes significant stochastic reconfiguration after each thermal cycle. We demonstrate that a single thermal cycle reshuffles the spectral landscape to an extent comparable to thousands of hours of continuous low-temperature evolution. Notably, no observable qubit degradation is attributed to the thermal cycling process itself. Overall, this study provides detailed insights into how thermal cycling affects superconducting quantum processors and underscores the necessity of adaptive recalibration protocols for scalable quantum systems.

\section{Experimental Setup}
We performed the characterization on a superconducting quantum processor with a flip-chip architecture. The device comprises two stacked chips fabricated on sapphire substrates: a top chip and a bottom chip. The top chip integrates 66 aluminum-based frequency-tunable transmon qubits together with 110 tunable couplers. The bottom chip houses the readout and control circuitry and is electrically connected to the top chip through indium bump bonds. Each qubit is controlled via a combined line for microwave drive and flux bias. Qubit state measurement is performed using dispersive readout through a resonator coupled to a Purcell filter. For this study, we selected a diverse subset of 27 qubits, labeled Q1 to Q27, for longitudinal stability analysis. The assembled chip is wire-bonded into a gold-plated copper package mounted on the mixing chamber of a dilution refrigerator. To shield the flux-tunable qubits from external magnetic noise, the sample holder is enclosed in a magnetic shielding cover. All input and output measurement lines are heavily attenuated and filtered at multiple thermal stages to suppress thermal and photon shot noise. During the entire characterization period, the base temperature of the system was maintained at approximately 20 mK.

With the device installed in this stable cryogenic environment, our systematic stability investigation spans approximately one year and encompasses four separate thermal cycles. Here, a thermal cycle refers to the process of warming the system to room temperature, above 290~K, and subsequently cooling it back to the base temperature. The longitudinal dataset tracks the evolution of the 27 qubits across these four thermal cycles of varying durations.

\section{Results and Discussion}
\begin{figure}[!htb]
\centering
\includegraphics[width=\columnwidth]{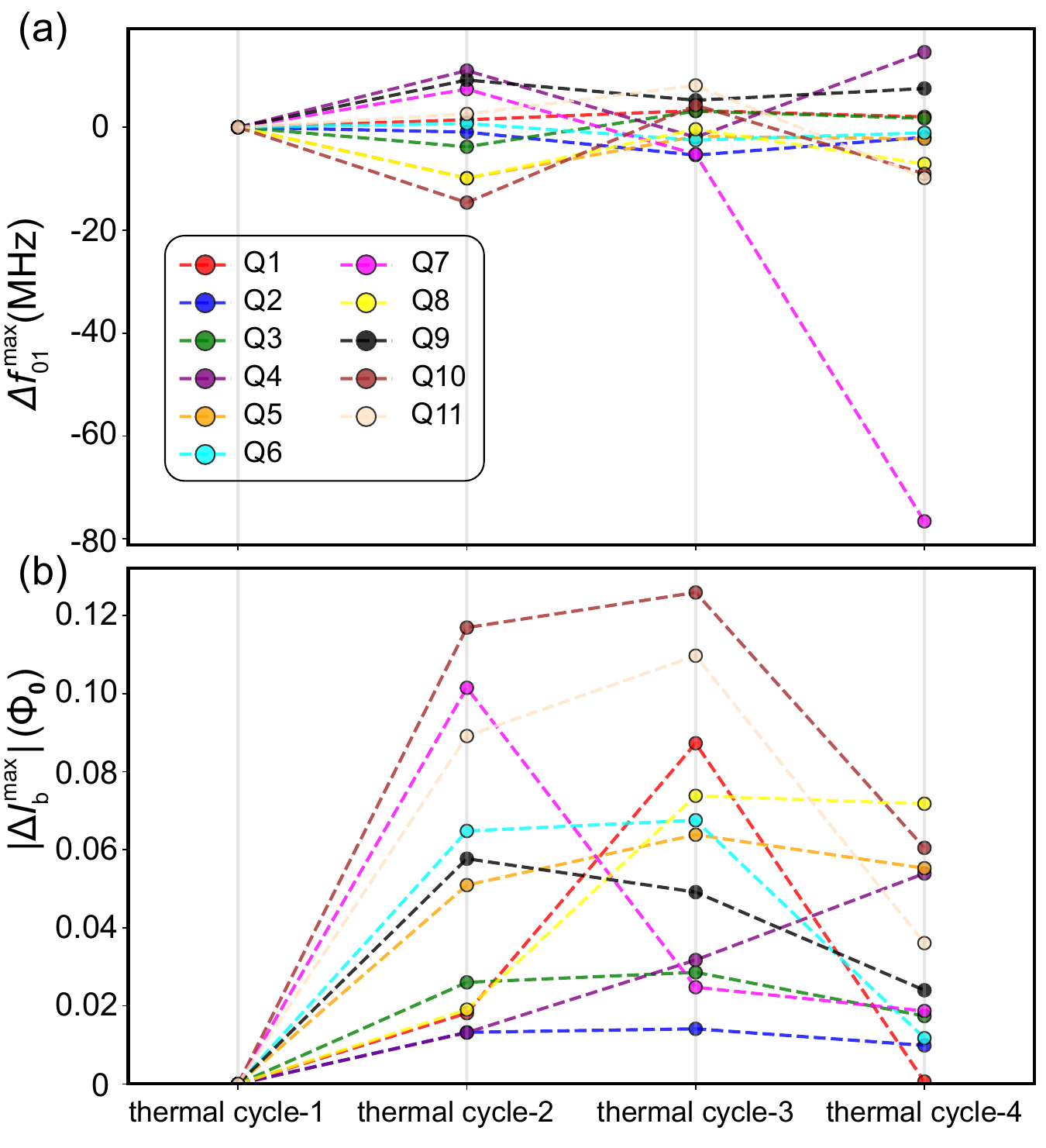}
\caption{Systematic characterization of parameter stability across four thermal cycles for 11 transmon qubits (Q1--Q11). (a) Evolution of the maximum qubit frequency deviation, $\Delta f_{01}^{\max}$. Values are relative to the thermal cycle-1 baseline. (b) Evolution of the absolute sweet-spot flux bias offset, $\Delta I_{b}^{\max}$. The colored dashed lines track individual qubits, revealing stochastic fluctuations.\label{fig:thermal_cycle_stability}}
\end{figure}

We begin by evaluating the stability of the static parameters across the four thermal cycles. For each cycle, we performed flux-dependent spectroscopy to characterize the qubit frequency response. Fitting the measured spectral arches allowed us to accurately determine the sweet-spot frequency ($f_{01}^{\max}$) and its corresponding flux bias offset ($I_b^{\max}$) for each qubit. The evolution of these parameters for a subset of 11 qubits (Q1--Q11) is depicted in Figure~\ref{fig:thermal_cycle_stability}, which displays the maximum qubit frequency ($f_{01}^{\max}$) and the magnitude of deviation in the sweet-spot flux bias offset ($|\Delta I_b^{\max}|$), both referenced to the cycle-1 baseline.

As shown in Figure~\ref{fig:thermal_cycle_stability}(a), $f_{01}^{\max}$ exhibits high stability. Excluding a single outlier (Q7 in cycle-4), the frequency deviations $|\Delta f_{01}^{\max}|$ for all qubits across all cycles remain confined within a narrow band of $\pm 20$~MHz. Given that the typical $f_{01}^{\max}$ for these symmetric-junction transmons is around $4.5$~GHz, this corresponds to a relative variation of less than $0.5\%$. The maximum frequency of a transmon qubit is governed by its Josephson energy $E_J$ and charging energy $E_C$, following $f_{01}^{\max} \approx \sqrt{8E_J E_C} - E_C$~\cite{george2017multiplexing, krantz2019quantum}. Consequently, variations in $f_{01}^{\max}$ serve as a reliable indicator of the structural integrity of the core quantum hardware, which encompasses both the lithographic capacitor geometry and the Josephson junction tunnel barriers. Given that thermal cycling is unlikely to affect geometric capacitance significantly, the observed stability of $f_{01}^{\max}$ suggests that the aluminum oxide tunnel barriers are highly robust and can withstand the substantial mechanical stresses induced by repeated thermal expansion and contraction.

In stark contrast to the stable qubit frequencies, the flux bias offsets exhibit significant stochastic fluctuations. The magnitude of these deviations, $|\Delta I_b^{\max}|$, for the same qubit subset is presented in Figure~\ref{fig:thermal_cycle_stability}(b). For instance, qubits Q10 and Q11 show pronounced deviations during cycle-2 and cycle-3, reaching offset magnitudes as large as approximately 0.12~$\Phi_0$. This fluctuating behavior of $I_b^{\max}$ is a characteristic signature of dynamic \textit{flux trapping} and subsequent \textit{reconfiguration} processes within the device~\cite{stan2004critical, song2009microwave}. Such behavior aligns with the view that thermal cycling can act as a reset mechanism for the magnetic environment of the chip.

\begin{figure}[!htb]
\centering
\includegraphics[width=\columnwidth]{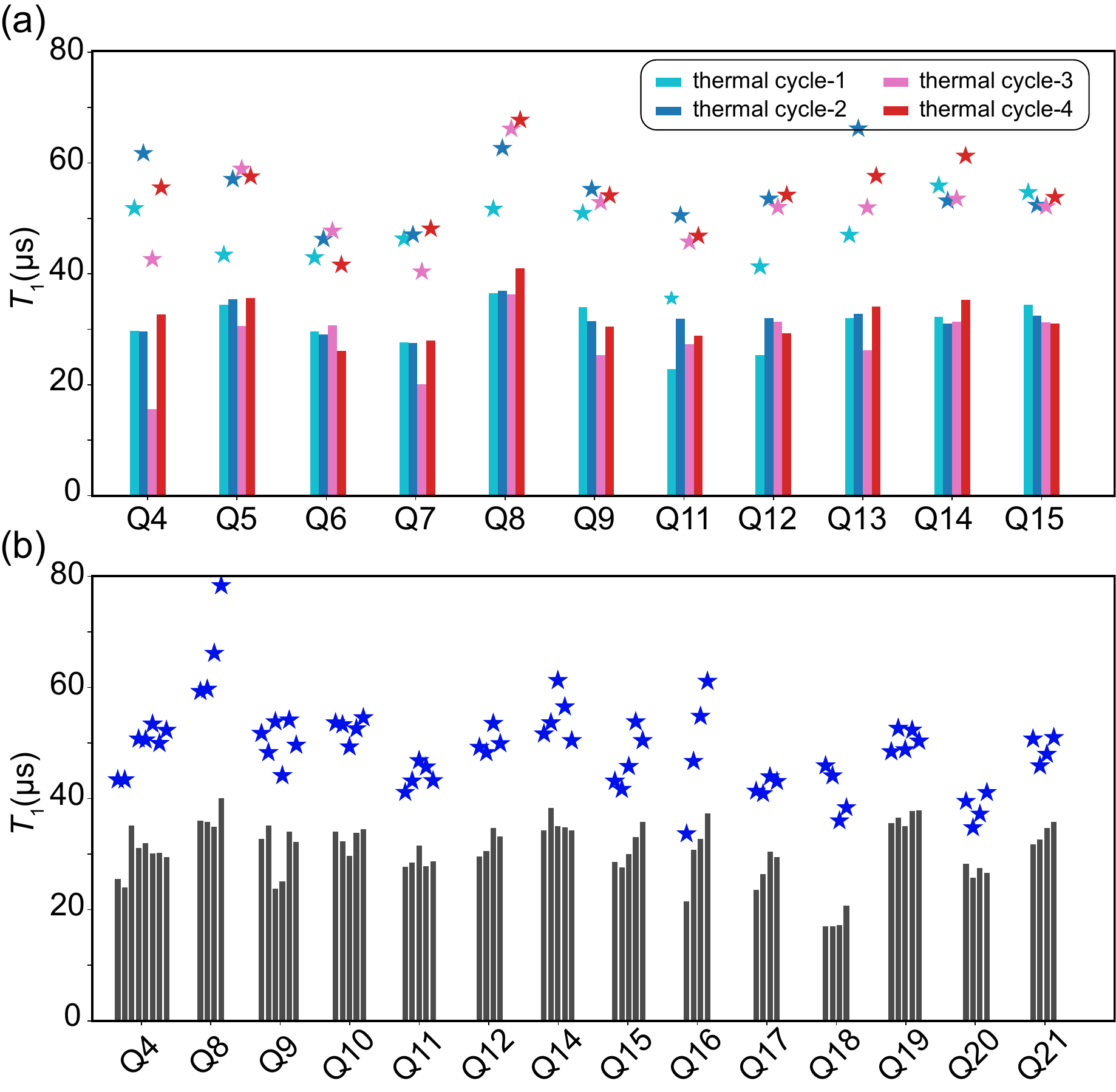}
\caption{Statistical characterization of energy relaxation time ($T_1$) stability. (a) Inter-cycle $T_1$ stability. For each qubit, the bar height represents the mean $T_1$ value calculated from multiple measurement sessions conducted within a single thermal cycle. The star marker ($*$) indicates the maximum $T_1$ value observed among those sessions in the same cycle. (b) Intra-cycle $T_1$ stability. For each qubit, the multiple bars represent individual measurement sessions taken sequentially within one thermal cycle. For each session (bar), its height corresponds to the mean $T_1$, and the associated star marker denotes the maximum $T_1$ measured within that specific session.\label{fig:T1_stability}}
\end{figure}

We then characterized the energy relaxation time, $T_1$ over four thermal cycles. In Figure~\ref{fig:T1_stability}, the $T_1$ evolution for 11 qubits across four thermal cycles is illustrated. Although thermal cycling can cause variations in qubit $T_1$, we find that it does not alter $T_1$ dramatically. For instance, high-coherence qubits like Q8 and Q14 maintain their average performance levels ($\sim 35\,\mu\mathrm{s}$ and $\sim 30\,\mu\mathrm{s}$, respectively) consistently across the intervening thermal cycles. This suggests that the primary loss channels, likely dominated by dielectric loss at the interfaces and in the bulk substrate~\cite{Wang2015, Dial2016,mcrae2020materials}, are structural in nature and remain unaffected by the thermal stress of warming and re-cooling. In contrast, the significant gap between the mean and maximum values highlights the impact of transient defects. Taking Q8 in thermal cycle-4 as an example, while its mean $T_1$ is $\sim 40\,\mu\mathrm{s}$, its maximum observed $T_1$ reaches nearly $70\,\mu\mathrm{s}$. The maximum values represent ``clean'' moments where the qubit frequency is temporarily detuned from strong TLSs, whereas the mean values include the penalty of random TLS coupling.

Notably, we observe a ``recovery'' behavior in several qubits (e.g., Q4, Q7, Q13, etc). These qubits exhibited suppressed mean $T_1$ values in a specific cycle (e.g., thermal cycle-3) but recovered to high-coherence levels in the subsequent thermal cycle-4. This reversibility strongly supports the hypothesis that inter-cycle variations are driven by transient environmental configurations rather than permanent material degradation.

To isolate the source of these fluctuations, we compare the inter-cycle data with intra-cycle statistics shown in Figure~\ref{fig:T1_stability}(b). Here, multiple measurement sessions were performed sequentially within a single representative thermal cycle. The temporal variance observed within this single thermal cycle is strikingly similar to the variance observed across different thermal cycles in Figure~\ref{fig:T1_stability}(a). Even within a stable cryogenic environment, qubits such as Q8 exhibit large fluctuations between the mean and maximum $T_1$ (with peak values reaching nearly $80\,\mu\mathrm{s}$). This result implies that the instability in coherence is not driven by the thermal cycling process itself, but rather by the continuous spectral diffusion of microscopic TLS defects~\cite{Klimov2018, Burnett2019}. These defects fluctuate stochastically in frequency over time, drifting in and out of resonance with the qubit~\cite{Muller2019}. The ``baseline'' quality of the device is preserved, and the observed instability is dominated by the intrinsic glassy dynamics of the TLS bath~\cite{Muller2019, Klimov2018}, which operates independently of the macroscopic thermal history.

In order to understand the microscopic origins of the coherence fluctuations, we move beyond single-point statistics to frequency-dependent relaxation spectroscopy. By stepping the qubit frequency $f_{01}$ and measuring the decay dynamics at each point, we generate a time-frequency spectrogram that acts as a unique ``fingerprint'' of the local defect environment. To quantify the stability of the qubit's microscopic environment against the fluctuating background of TLS defects, we analyze the reproducibility of these frequency-resolved measurements. We treat the raw time-frequency data as a matrix, $\mathbf{\Phi}$, with dimensions $N_{\tau} \times N_{\omega}$. Here, each element $\Phi_{ij}$ represents the qubit population $P(\omega_j, \tau_i)$, mapping the interaction strength between the qubit and its local defect bath across both the frequency ($\omega_{01}$) and time ($\tau$) domains.

For the comparative analysis, we evaluate two spectral snapshots, $\mathbf{\Phi}_A$ and $\mathbf{\Phi}_B$, obtained either from the same thermal cycle (intra-cycle stability) or distinct thermal cycles (inter-cycle reproducibility). Since the raw population probabilities are susceptible to systematic drifts in readout fidelity or thermal equilibrium populations, direct subtraction is unreliable. To address this, we project the raw decay maps into a standardized feature space via Z-score normalization. In this normalization, the elements of the renormalized matrix $\mathbf{X}'$ (where $\mathbf{X} \in \{\mathbf{\Phi}_A, \mathbf{\Phi}_B\}$) are given by:

\begin{equation}
    X'_{ij} = \frac{X_{ij} - \langle \mathbf{X} \rangle}{\sigma_{\mathbf{X}}},
    \label{eq:normalization}
\end{equation}
where $\langle \mathbf{X} \rangle$ and $\sigma_{\mathbf{X}}$ represent the global mean and standard deviation of the map, respectively. This transformation effectively decouples the structural topography of the decay features (e.g., TLS-induced relaxation streaks) from global variations in measurement contrast.

\begin{figure}[!htb]
\centering
\includegraphics[width=\columnwidth]{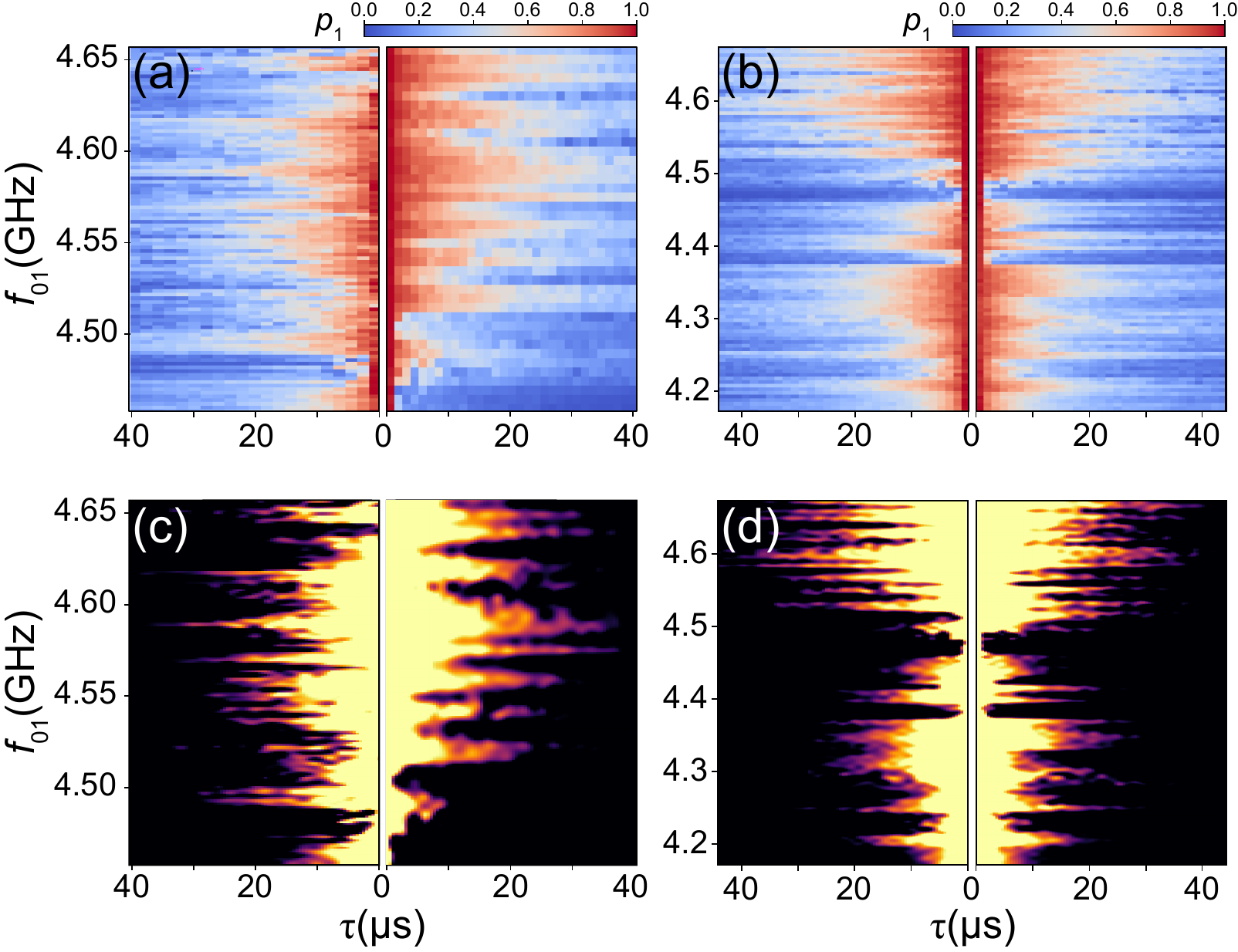}
\caption{Comparison of frequency-dependent energy relaxation ($T_1$) landscapes. (a) Raw time-frequency spectrograms of the excited state probability $p_1$ comparing data from thermal cycle-4 (left panel) and thermal cycle-3 (right panel). The vertical axis represents the qubit frequency $f_{01}$, and the horizontal axis represents the delay time $\tau$. (b) Raw time-frequency spectrograms of $p_1$ from two independent measurements within thermal cycle-3. (c) Z-score normalized $p_1$ maps corresponding to (a). (d) Z-score normalized $p_1$ maps corresponding to (b).\label{fig:tls_fingerprint}}
\end{figure}

The application of this framework is visualized in Figure~\ref{fig:tls_fingerprint}. Raw spectrograms comparing data from thermal cycle-4 (left) and thermal cycle-3 (right) are provided in Figure~\ref{fig:tls_fingerprint}(a), revealing uncorrelated TLS defect positions that indicate a redistribution between cycles. In contrast, as shown in Figure~\ref{fig:tls_fingerprint}(b), two independent measurements within thermal cycle-3 reveal defect signatures at consistent frequencies, confirming spectral stability within a single cycle. The normalized maps in Figure~\ref{fig:tls_fingerprint}(c) (corresponding to panel a) and (d) (corresponding to panel b) amplify these differences. As observed in Figure~\ref{fig:tls_fingerprint}(c), the normalized maps display no structural correlation between cycles, whereas the mirror-symmetric patterns in Figure~\ref{fig:tls_fingerprint}(d) confirm high stability within a single cycle. This comparison clearly confirms that thermal cycling significantly reconfigures the TLS environment between cycles, whereas the spectral fingerprint remains highly stable within a single thermal cycle.

The structural dissimilarity is captured by the Euclidean distance between the feature matrices. In our experimental protocol, the sampling dimensions ($N_{\tau}, N_{\omega}$) are kept consistent across all measurement sessions to ensure that the calculated divergence is intrinsic to the spectral landscape rather than an artifact of resolution disparity. Accordingly, we define the \textit{normalized spectral divergence}, $\delta$, as:
\begin{equation}
    \delta = \frac{1}{N_{\tau} N_{\omega}} \sqrt{ \sum_{i=1}^{N_{\tau}}\sum_{j=1}^{N_{\omega}} \left( \Phi'_{A,ij} - \Phi'_{B,ij} \right)^2 }.
    \label{eq:divergence_density}
\end{equation}
This quantity reflects the average contribution of each pixel to the total Euclidean distance in the standardized decay landscape. We subsequently map this divergence to the \textit{$T_1$ Spectral Topography Fidelity (STF)}, denoted as $\rho$, defined as:
\begin{equation}
    \rho  = \frac{\alpha}{\delta}.
    \label{eq:similarity_score}
\end{equation}
Here, $\alpha$ is an empirical scaling factor determined to map the intrinsic noise floor of our measurement setup to a reference scale. Consequently, $\rho$ serves as a robust score of environmental congruence: high $\rho$ values denote a highly reproducible TLS distribution, while sharp drops in $\rho$ quantitatively reveal the reconfiguration of the spectral defect landscape triggered by thermal cycling.
\begin{figure*}[ht]
\centering
\includegraphics[width=0.95\textwidth]{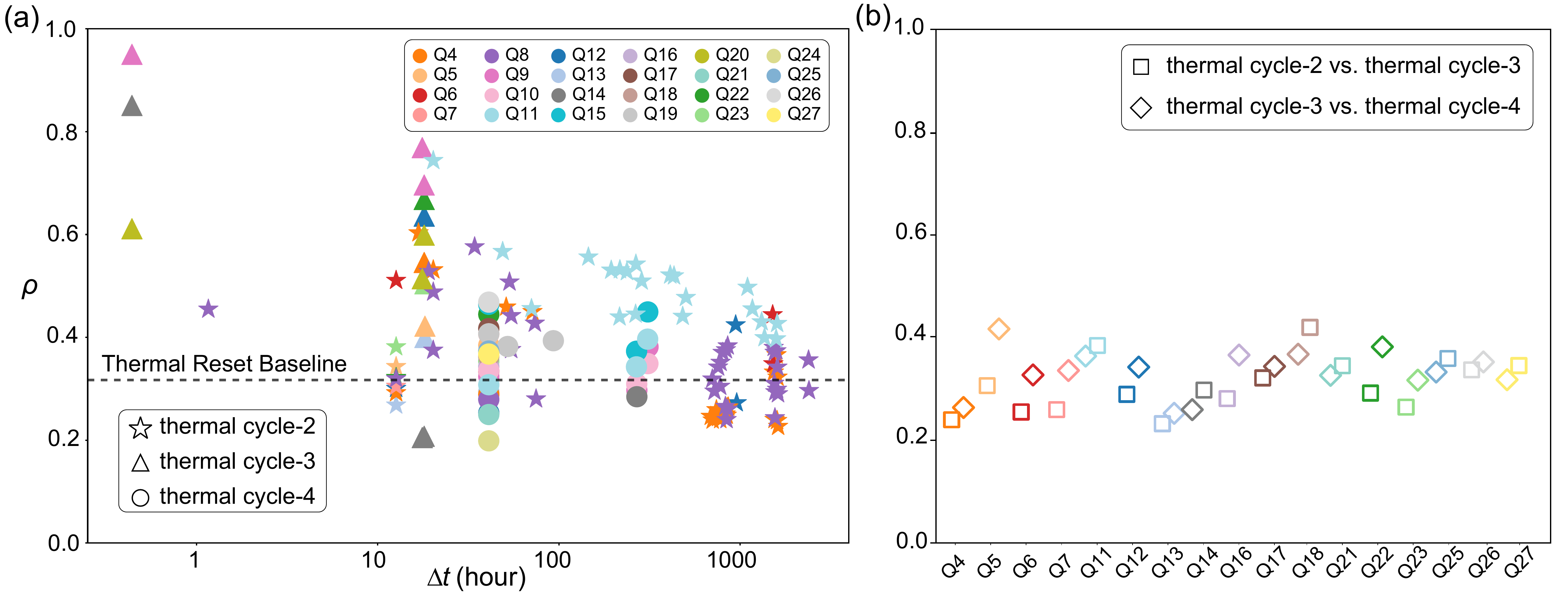}
\caption{Quantitative analysis of spectral stability using the $T_1$ STF, $\rho$. (a) Intra-thermal cycle temporal evolution of $\rho$. $\Delta t$ represents the time interval between two measurement sessions within the same thermal cycle. The markers distinguish the datasets from thermal cycle-2 (stars), thermal cycle-3 (triangles), and thermal cycle-4 (circles), while colors correspond to different qubits. The horizontal dashed gray line marks the mean fidelity level of the inter-thermal cycle data (from panel b), representing the baseline of a randomized environment. (b) Inter-thermal cycle reproducibility. The scatter plot displays the $T_1$-STF calculated between maps of the same qubit taken from adjacent thermal cycles (thermal cycle-2 vs. 3, and thermal cycle-3 vs. 4). The consistently low values ($\rho \sim 0.3$) indicate that the microscopic defect landscape is effectively uncorrelated after a thermal cycle.\label{fig:stf_analysis}}
\end{figure*}
We then apply the previously defined $T_1$-STF metric to the full dataset to quantify the timescales of environmental memory. The evolution of the $T_1$-STF within a single thermal cycle is displayed in Figure~\ref{fig:stf_analysis}(a). The data reveal a clear decaying trend that is approximately linear on a logarithmic time scale, which is characteristic of spectral diffusion in amorphous solids~\cite{Black1977}. Over time, slower thermal fluctuators change their states. This progressively modifies the local dielectric environment and causes the qubit's spectral fingerprint to drift~\cite{Klimov2018}. Consequently, even within a stable millikelvin environment, the memory of the exact defect configuration gradually fades over hundreds of hours, with $\rho$ decreasing from above $0.8$ to approximately $0.4$.

We extend this analysis to comparisons across different thermal cycles, as presented in Figure~\ref{fig:stf_analysis}(b). Here we evaluate the $T_1$-STF between thermal cycle~2 and~3, and between cycle~3 and~4. The consistently low values, $\rho \sim 0.3$, indicate that the microscopic defect landscape becomes effectively uncorrelated after a thermal cycle. Notably, this baseline level matches the fidelity observed after more than $1000$~hours of continuous evolution in Figure~\ref{fig:stf_analysis}(a).

The comparison between intra-cycle and inter-cycle behaviour provides a key physical insight. A single thermal cycle randomises the system sufficiently to bypass the slow temporal diffusion process. It supplies enough thermal energy to overcome the high energy barriers in the TLS configuration space. This achieves an environmental reconfiguration that, according to the observed logarithmic decay trend, would theoretically require months or years to occur via tunnelling-induced diffusion at cryogenic temperatures. Therefore, thermal cycling acts as a stochastic hard reset, effectively erasing the specific microscopic correlations established during the previous cycle. This also implies that even within the same thermal cycle, the overall TLS background can change dramatically. Such changes may arise from interactions among TLSs~\cite{muller2015, faoro2015}, drifting quasiparticle traps~\cite{deGraaf2020}, or ionisation-induced TLS reconfiguration~\cite{Thorbeck2023}.

\section{Conclusions}
Our longitudinal study of 27 transmon qubits over nine months and four thermal cycles reveals a clear stability hierarchy for superconducting quantum hardware. The intrinsic parameters that define qubit frequency (\(f_{01}\)) show strong robustness, with typical variations remaining within \(\pm 0.5\%\). Furthermore, the baseline energy relaxation time (\(T_1\)) exhibits no signs of degradation, which confirms the structural integrity of the core fabricated elements. In contrast, the environmental variables undergo significant stochastic reconfiguration after each thermal cycle. These variables include the background magnetic flux and the microscopic distribution of two-level system defects. Through quantitative analysis using the \(T_1\) Spectral Topography Fidelity metric, we demonstrate that a single thermal cycle functions as a stochastic hard reset. It scrambles the spectral landscape to a degree that would otherwise require thousands of hours of continuous cryogenic diffusion. These results have direct practical implications. They affirm the reliability of current fabrication techniques for long-term operation, yet they also indicate that fully automated recalibration protocols are essential to manage the environmental randomness introduced after each thermal cycle in large-scale systems. Additionally, controlled thermal cycling emerges as a potential strategy to probabilistically reset problematic defect configurations. By bridging microscopic defect dynamics to system-level quantum engineering, this work provides a foundation for the adaptive calibration and maintenance necessary for scalable quantum computing.

\section*{Funding}
This research was funded by the National Key R\&D Program of China (Grant No. 2024YFB4504000), the Fundamental and Interdisciplinary Disciplines Breakthrough Plan of the Ministry of Education of China (Grant No. JYB2025XDXM202) and the Aid Program for Science and Technology Innovative Research Team in Higher Educational Institutions of Hunan Province.

\bibliographystyle{IEEEtran}
\bibliography{Bibfile.bib}

\end{document}